\documentclass{PoS}
\usepackage{upgreek}
\title{The first GCT camera for the Cherenkov Telescope Array.}

\ShortTitle{GCT camera.}

\author{\speaker{A. De Franco}$^a$\footnote{Supported by the EU FP7-PEOPLE-2012-ITN project nr 317446, INFIERI, ``Intelligent Fast Interconnected and Efficient Devices for Frontier Exploitation in Research and Industry``.}, R. White$^b$, D. Allan$^c$, T. Armstrong$^c$, T. Ashton$^g$, A. Balzer$^d$, D.~Berge$^d$, R. Bose$^{h}$, A.~M. Brown$^c$, J. Buckley$^{h}$, P.~M. Chadwick$^c$, P. Cooke$^{e}$, G. Cotter$^a$, M.~K. Daniel$^{e}$, S. Funk$^{f}$, T. Greenshaw$^{e}$, J. Hinton$^b$, M. Kraus$^{f}$, J. Lapington$^g$, P. Molyneux$^g$, P. Moore$^{h}$, S. Nolan$^{c}$, A. Okumura$^{i, b}$, D. Ross$^g$, C. Rulten$^{j}$, J. Schmoll$^{c}$, H. Schoorlemmer$^b$, M. Stephan$^d$, P. Sutcliffe$^{e}$, H. Tajima$^{i}$,  J. Thornhill$^{g}$, L. Tibaldo$^{k}$, G. Varner$^{l}$, J. Watson$^a$, A. Zink$^{f}$ for the CTA Consortium\footnote{Full consortium author list at http://cta-observatory.org}\\
        E-mail: \email{andrea.defranco@physics.ox.ac.uk}

{\footnotesize
$^{a}$ University of Oxford, Department of Astrophysics, Oxford, OX1 3RH, UK;\\
$^{b}$ Max-Planck-Institut f{\"u}r Kernphysik, Saupfercheckweg 1, 69117 Heidelberg, Germany;\\
$^{c}$ University of Durham, Department of Physics, Durham, DH1 3LE, UK;\\
$^{d}$ GRAPPA, Anton Pannekoek Institute for Astronomy, University of Amsterdam,  Science Park 904, 1098 XH Amsterdam, The Netherlands;\\ 
$^{e}$ University of Liverpool, Department of Physics, Liverpool, L69 7ZE, UK;\\
$^{f}$ Physikalisches Institut der Friedrich-Alexander, Universit{\"a}t Erlangen-N{\"u}rnberg, Erwin-Rommel-Str. 1, B{\"u}ro 219, D-91058, Erlangen, Germany;\\
$^{g}$ University of Leicester, Department of Physics and Astronomy, Leicester, LE1 7RH, UK;\\
$^{h}$ Washington University at St Louis,1 Brookings Dr, St. Louis, MO 63130, United States;\\
$^{i}$ Solar-Terrestrial Environment Laboratory, Nagoya University, Furo-cho, Chikusa-ku, Nagoya, Japan;\\
$^{j}$ School of Physics and Astronomy, 116 Church Street Southeast, Minneapolis, MN 55455, United States;\\
$^{k}$ SLAC, KIPAC, 2575 Sand Hill Road, M/S 29 Menlo Park, CA  94025, United States;\\
$^{l}$ Department of Physics and Astronomy, University of Hawaii, 2505 Correa Road, Honolulu, HI 96822, USA}
}	

\abstract{The Gamma Cherenkov Telescope (GCT) is proposed to be part of the Small Size Telescope (SST) array of the Cherenkov Telescope Array (CTA).  The GCT dual-mirror optical design allows the use of a compact camera of diameter roughly 0.4~m. The curved focal plane is equipped with 2048 pixels of $\sim$0.2$^{\circ}$ angular size, resulting in a field of view of $\sim$9$^{\circ}$.  The GCT camera is designed to record the flashes of Cherenkov light from electromagnetic cascades, which last only a few tens of nanoseconds. Modules based on custom ASICs provide the required fast electronics, facilitating sampling and digitisation as well as first level of triggering. The first GCT camera prototype is currently being commissioned in the UK. On-telescope tests are planned later this year. Here we give a detailed description of the camera prototype and present recent progress with testing and commissioning.}

\FullConference{The 34th International Cosmic Ray Conference,\\
  30 July- 6 August, 2015\\
  The Hague, The Netherlands}

\begin{document}

\section{Introduction}
The Southern hemisphere site of the Cherenkov Telescope Array (CTA) \cite{CTA} is envisaged to host $\sim$70 Small-Sized Telescopes (SST) \cite{SST}, optimised for the highest gamma-ray energies that CTA will be sensitive to. The Gamma-ray Cherenkov Telescope (GCT), a sub-consortium of CTA, proposes to construct 35 SSTs, designed to cover the energy range from about 1 to 300 TeV. The GCT telescope is a dual mirror Schwarzschild-Couder design with a primary mirror of diameter 4~m, a secondary mirror of diameter 2~m, a focal length of 2.3~m and a focal ratio of 0.57 (see \cite{GCT_Telescope}) resulting in a spherical focal plane with a radius of curvature 1.0~m. The GCT camera is designed to record flashes of Cherenkov light lasting from a few to a few tens of nanoseconds, with typical image width and length of $\sim$0.2$^{\circ} \times $1.0$^{\circ}$ and promises a low-cost, high-reliability, high-data-quality solution for a dual-mirror SST. The small focal length of the telescope implies that an approximately 0.2$^{\circ}$ angular pixel size (as required by CTA for the SSTs) is achievable with pixels of physical dimensions of 6 to 7 mm, while the dual-mirror optics ensure that the point spread function (PSF) of the telescope is below 6~mm up to field angles of 4.5$^{\circ}$.  A field of view (FoV) of 8$^{\circ}$ (again required by CTA for the SSTs) can therefore be covered with a camera of diameter about 0.4 m, composed of 2048 pixels. This allows the use of commercially available photosensor arrays, significantly reducing the complexity and cost of the camera. Suitable packages consisting of multi-anode photomultipliers (MAPMs) or silicon photomultipliers (SiPMs) are under investigation for the GCT via the development of two prototypes (CHEC-M) and (CHEC-S) respectively, where CHEC stands for Compact High Energy Camera. In this paper we focus on the former which is now under testing in the lab. In section \ref{GCT-SiPM} we briefly report the ongoing work towards the completion of the second camera.

\section{CHEC-M Design}

\begin{figure}[!ht]
  \centering
\resizebox{1\columnwidth}{!}{\includegraphics[trim=1cm 0cm 1cm 0cm, clip=true]{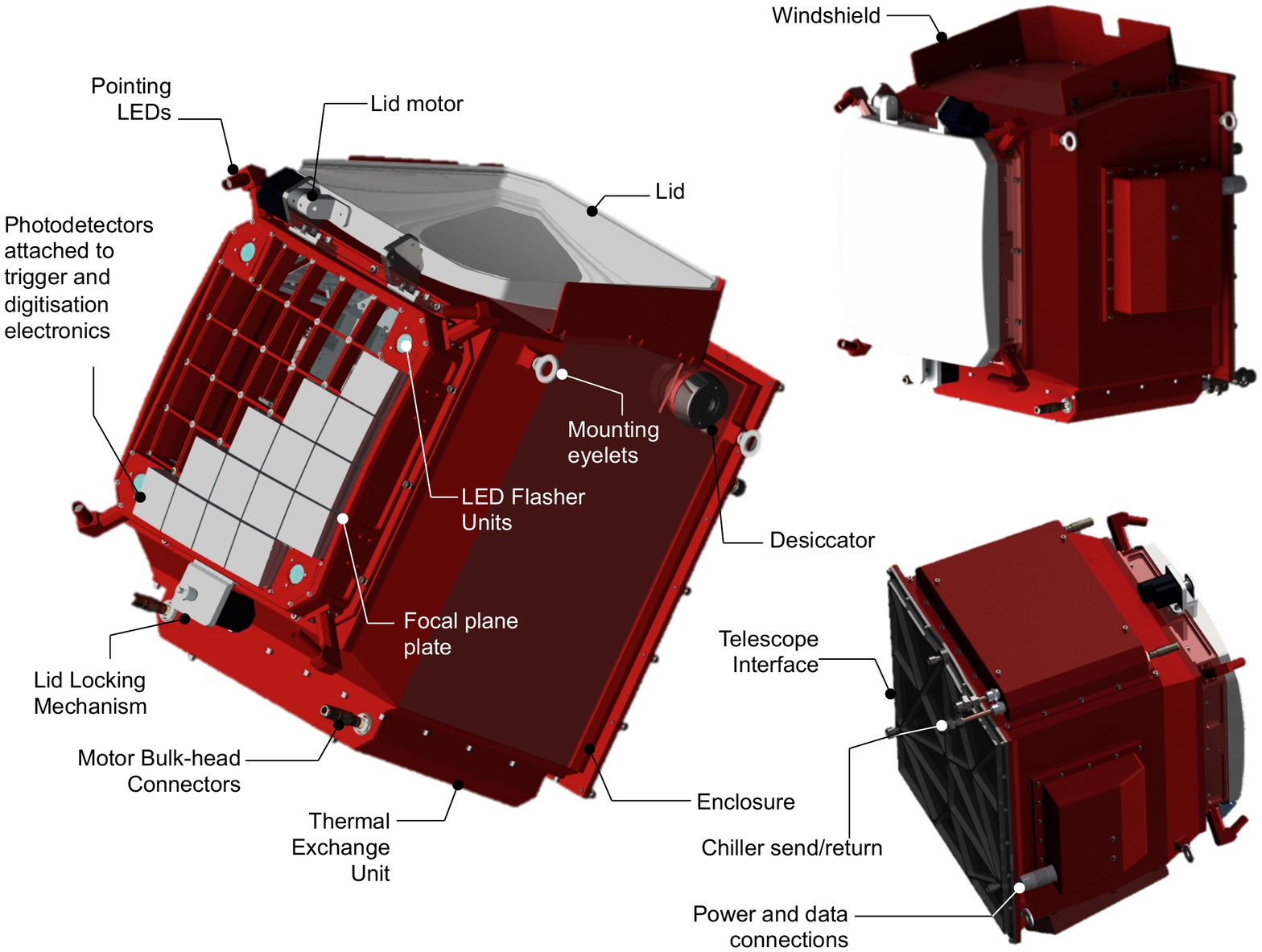}}
\caption{A detailed overview of the GCT camera. The CAD model shown is
  for the prototype camera equipped with MAPMs (CHEC-M). In addition a protective window
  may be used to protect the photodetectors.}
  \label{FigCam}
\end{figure}

\begin{figure}[!ht]
  \centering
  \makebox[1\columnwidth][c]{\includegraphics[trim=3.5cm 1.5cm 4.5cm 0cm,
    clip=true, width=1.16\textwidth]{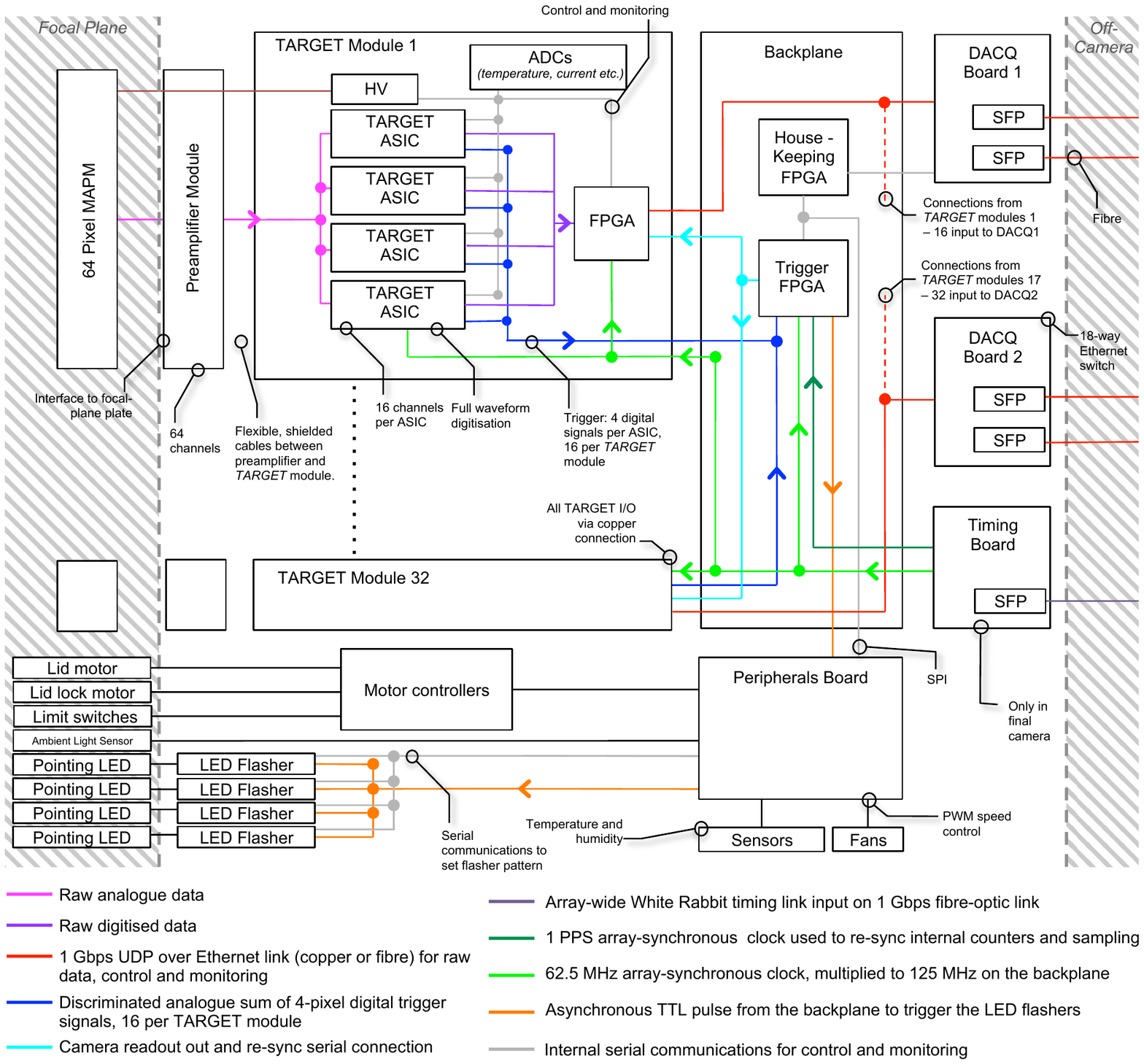}}
  \caption{A schematic showing the logical elements of the GCT camera,
    the communication between those elements, the raw data flow
    through the camera, the trigger architecture and the clock
    distribution scheme. Power distribution is excluded for
    simplicity.}
  \label{FigElectronics}
\end{figure}

A CAD image of CHEC-M is shown in Figure~\ref{FigCam} showing the principle external components, whilst Figure \ref{FigElectronics} shows the internal electronics architecture for the camera. The camera is designed to provide full waveform information for every pixel in every event. It is based on 32 camera modules each containing a 64-channel MAPM attached via a preamplifier board to a front-end module called TARGET providing digitisation and triggering. The modules are mounted in an internal rack and directly to a backplane PCB, which in-turn routes raw data to two data acquisition boards (DACQ). The position of the MAPMs at the focal plane is defined and aligned on the focal plane with the required accuracy of better than 0.35 ~mm. Individually shielded ribbon cables between the preamplifiers and TARGET modules are used to match the different physical geometries of the internal rack and the curved focal plane. LED flasher units placed in the corner of the camera provide calibration over a range of illumination intensities via reflection from the secondary mirror (see \cite{Flasher} for further details). Internal fans circulate air via a set of mechanical baffles. An external lid system provides protection from the elements. The lid, fans, LED flashers and a set of internal sensors (temperature and humidity) are controlled via a peripherals board. The camera is liquid cooled via a chiller unit mounted on the telescope. Camera readout and communications take place via optical fibre. The camera is powered by a single 12~V DC supply, mounted at the rear of the secondary mirror, and the power consumption is $\sim$450~W.

\subsection{Front-End Electronics}

\begin{figure}[!ht]
\begin{centering}
\resizebox{1\columnwidth}{!}{\includegraphics[trim=3cm 3cm 3cm 2cm, clip=true]{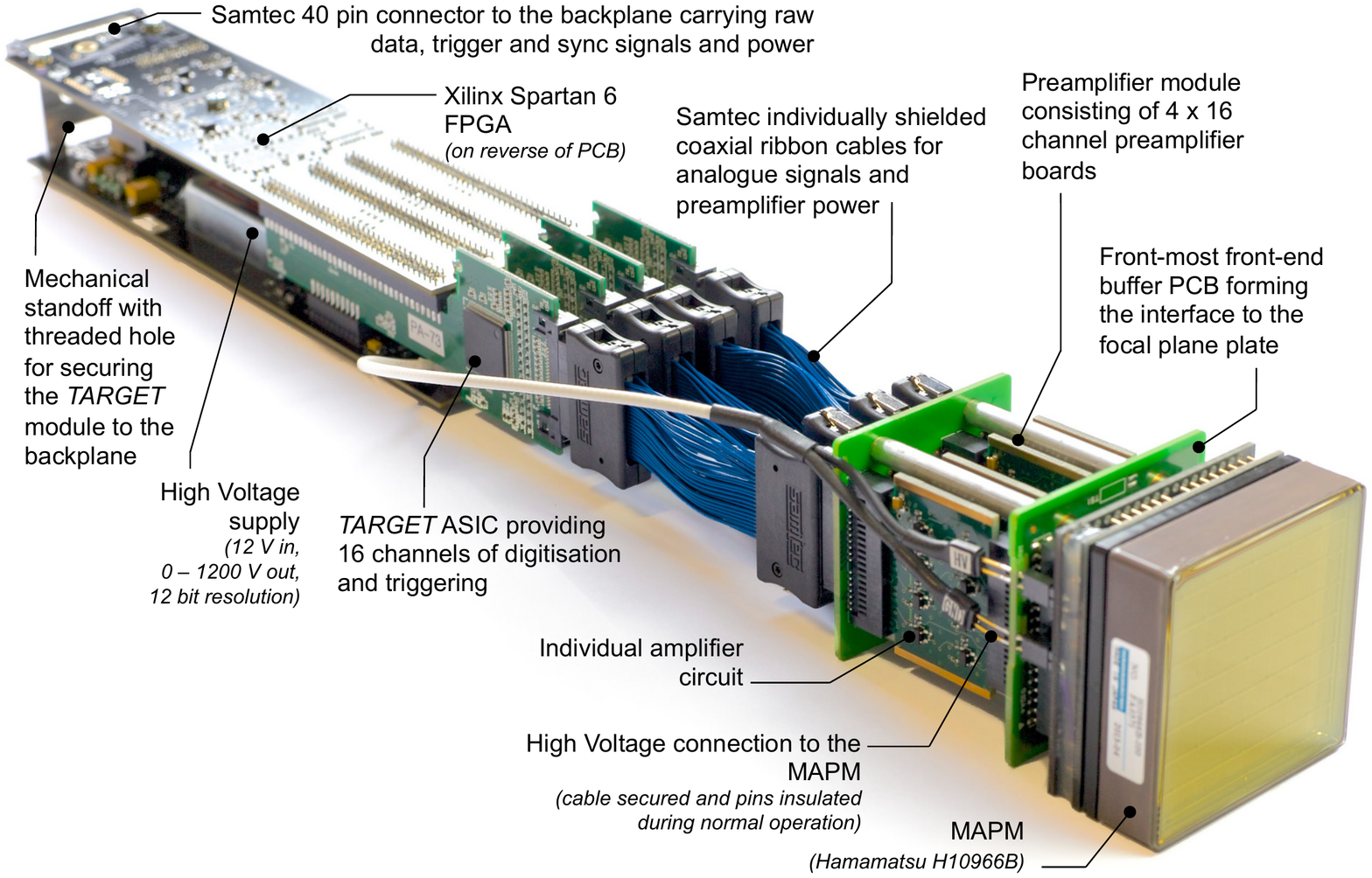}} 
\caption{Photograph of a TARGET 5 based front-end camera module used in CHEC-M.}
\label{FigTM}
\end{centering}
\end{figure}

Figure~\ref{FigTM} shows the front-end camera module used in CHEC-M. Each MAPM has 64 pixels with physical size of $\sim$6~mm$\times$6~mm. The photosensor is connected to a preamplifier module, containing 4 PCBs of 16 channels each, which shape the signal from the fast detector stretching the signal to allow for an optimal coincidence time window for the trigger. This optimal shape is determined from Monte Carlo simulations to be: 5.5--10.5~ns FWHM, 3.5--6.0~ns rise time. The shaped pulses are then fed to the TARGET module via ribbon cables. The TARGET module is based on four custom TARGET ASICs \cite{TARGET}, each one of which accepts 16 analogue inputs from the preamplifier module. The shaped signal is digitised with 12-bit resolution at 1~GS/s over a programmable read out window in 32~ns steps (96~ns is currently envisaged for normal operation). The conversion and readout lead to a dead time of $\sim$20 $\upmu$s, which is acceptable at the maximum average camera trigger rate of $\sim$600~Hz. The CHEC-M TARGET modules use TARGET 5 ASICs which show an RMS noise of 0.5~mV and a dynamic range of 1.2~V. Resulting in a pixel saturation at $\sim$500 p.e. assuming operation at a gain level such as 1~p.e. corresponds to 5 times the noise RMS. Simulations have shown that the waveform information can be used to extend the dynamic range beyond saturation. The TARGET module also provides the first level of trigger based on the discrimination of the analog sum of 4 neighbouring pixels (super-pixel). The TARGET module contains a Xilinx Spartan-6 FPGA to process trigger signals (dark and light blue lines in Fig. \ref{FigElectronics}) and read waveforms(gray line in Fig. \ref{FigElectronics}) from arbitrary storage capacitors. Furthermore it provide interface for slow control: set the supplied HV to the MAPM (operating range 800--1100 V), monitor status of the module, configure and read out the ASICs. Serial data is then output from the TARGET module over ethernet using UDP protocol for raw data, control and monitoring.

\subsection{Back-End Electronics}
A single backplane PCB is used to gather the signals from all the 32 TARGET modules. The backplane routes the UDP raw data lines to the DACQ boards, distributes a synchronous clock signal, supplies the TARGET modules with power (and control of that power), and provides the camera-level trigger. A Xilinx Virtex 6 FPGA accepts all 512 LVDS first-level trigger lines from the TARGET modules and implements a camera-level trigger algorithm (currently any two triggered neighbouring super pixel patches trigger the camera). Upon a valid trigger the TARGET modules are read out and a unique event number is added to the raw data. In the final CTA-scheme a timing board based on the White Rabbit protocol\cite{WhiteRabbit} will be used to provide an absolute timestamp and associate it with the unique event number. For CHEC-M the DACQ boards are capable of implementing a White Rabbit timing interface on-board. All raw data, control and monitoring communication to the camera is via four 1~Gbps fibre optic cables attached to the DACQ boards.

\section{CHEC-M Commissioning}
\begin{figure}
\begin{centering}
\resizebox{1\columnwidth}{!}{\includegraphics[trim=2cm 6.5cm 3cm 0cm, clip=true]{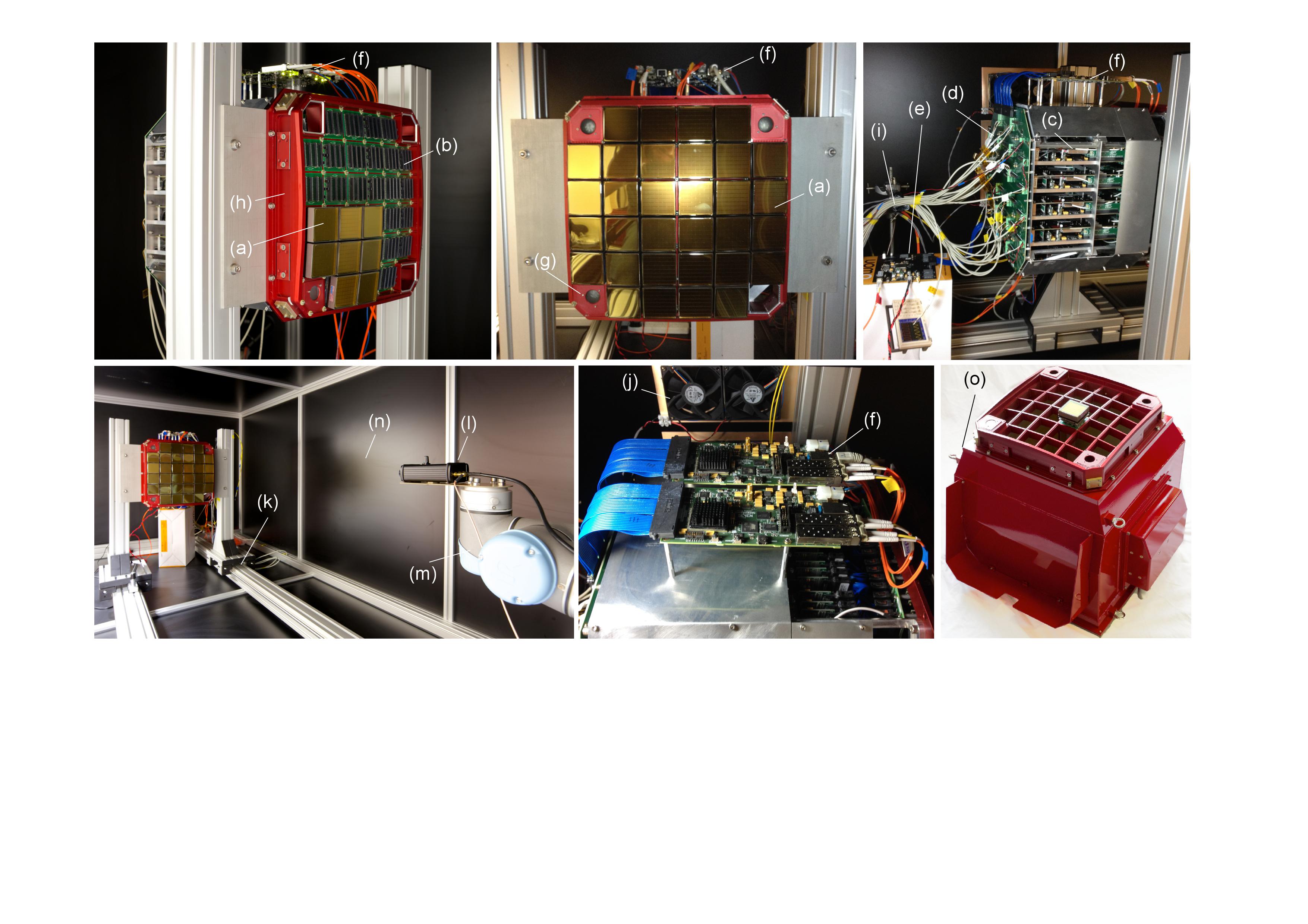}} 
\caption{CHEC-M: (a) MAPM photodetectors, (b) preamplifier modules, (c)
  TARGET modules, (d) proto-Backplane, (e) FPGA test board, (f) DACQ
  boards, (g) calibration flashers, (h) internal mechanics consisting
  of focal plane plate and electronics rack, (i) 12 V input cables (j)
  lab fan-cooling, (k) sliding rail, (l) illumination source, (m) robot arm, (n) dark box, (o) external mechanical enclosure.}
\label{FigChecM}
\end{centering}
\end{figure}

Figure~\ref{FigChecM} shows the fully assembled prototype camera populated with all front-end modules, internal mechanics and DACQ boards. A prototype backplane PCB is used to verify camera readout whilst the real backplane is undergoing separate lab-tests and firmware development. The camera is held horizontally on a sliding rail in a dark box also hosting a robot arm holding a near UV laser as shown in Figure~\ref{FigChecM}. For the purpose of flat fielding a diffuser is mounted in front of the laser and the robot arm is held as far from the camera as the physical size of the dark box allows to have the most uniform light beam possible.

\begin{figure}
\begin{centering}
\resizebox{0.9\columnwidth}{!}{\includegraphics[trim=2.2cm 7cm 2.5cm 1cm,
  clip=true]{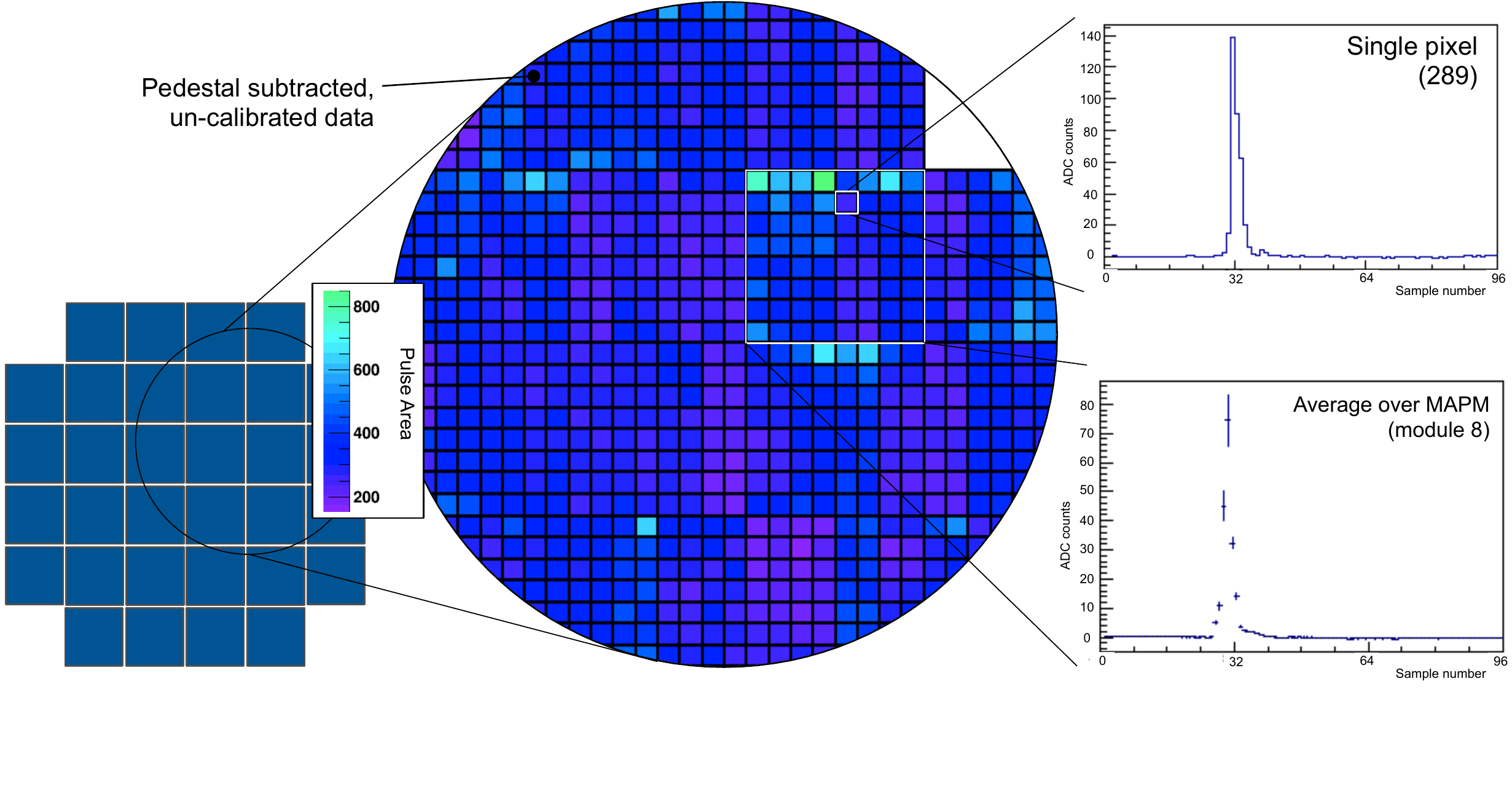}}
\resizebox{1.0\columnwidth}{!}{\includegraphics[trim=1cm 11cm 2cm 1cm, clip=true]{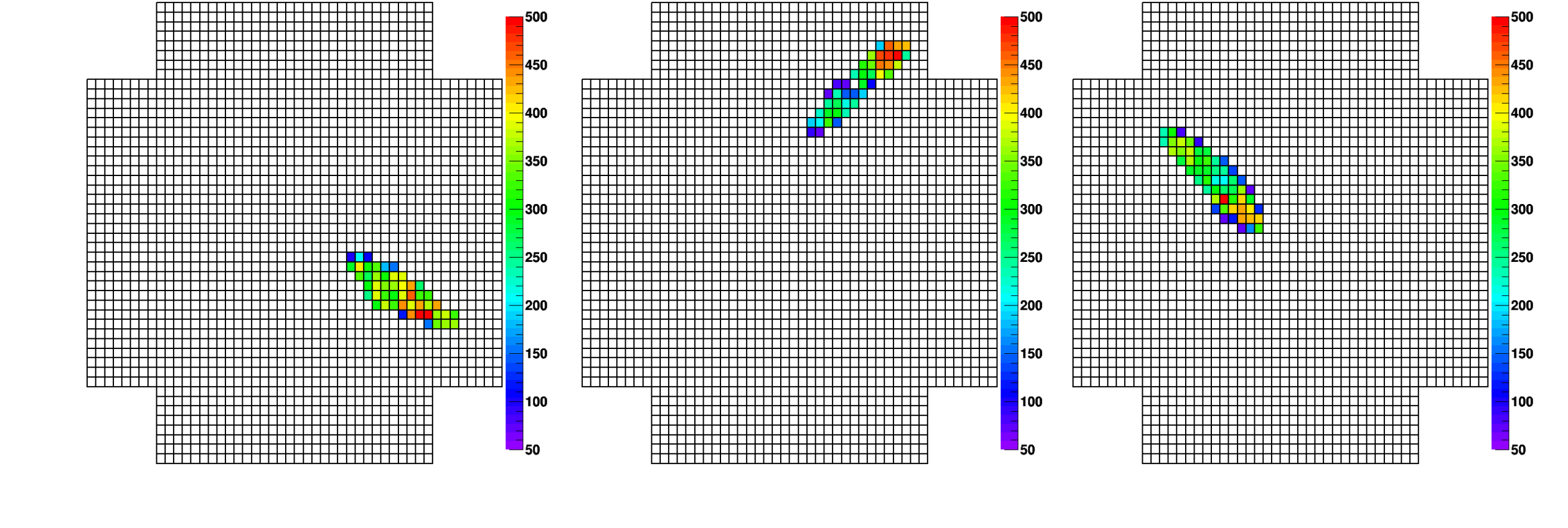}}
\caption{Initial commissioning results from CHEC-M measuring uniform light flashes. (top, middle) pulse
  area, (top, right, top) example waveforms for an individual
  pixel and (top, right, bottom)  averaged of an MAPM. (bottom) images taken with a mask in-front of the camera
  to approximate the shape of Chernekov light flashes from gamma-ray initiated showers.}
\label{FigCom1}
\end{centering}
\end{figure}

Calibration and refining of the data-taking process is in progress. Figure~\ref{FigCom1} shows some initial commissioning results from CHEC-M, showing the pulse area as measured from an approximately uniform light flash (top, middle) and example waveforms from the same data for an individual pixel (top, right, bottom) and averaged over an MAPM (top, right, top). The bottom images are taken with a mask in front of the camera to approximate a Cherenkov image with a cut on the pulse peak height. In both cases, minimal gain-matching between MAPMs has been applied and pedestals have been subtracted, but no other calibration has been applied.  

%Figure~\ref{FigCom1} shows the arrival time of laser pulses in the camera. A spread of XXX~ns is measured, consistent with the 1~GS/s sampling rate and within the CTA requirements XXX~ns. 

Once commissioned, testing will include illuminating the full camera with uniform flashes of light at varying brightness (using both a laser and a CHEC LED flasher) in the presence of a realistic background light level (from a diffuse white-light source). The robot arm will also allow the focal plane to be scanned with a narrow beam from the laser to measure the variation in response as the edge of pixels and gaps between MAPMs are encountered as well as the angular response of the pixels.  

The GCT camera has to operate reliable in a very harsh environment for more than 15 years with only minor maintenance. Therefore an empty enclosure was artificially exposed in the lab to verify performance against CTA requirement. The tests performed were: impact testing, water ingress, temperature cycling (followed by further water-ingress testing), UV exposure and wind testing. Wind-tunnel tests showed we can operate the lid up to a maximum wind load of 50~km/hour.  Impacts on both the camera body and lid showed no ill effects, the camera body (including the interface with the faceplate) showed no evidence of water ingress, and there was no damage during temperature cycling between ${-}25$ and +40$^\circ$C or during UV exposure. Water ingress was found to be present via the seal between the camera lid and body. This lead to a minor, but important redesign to ensure camera survival on site.

On-telescope tests are planned to take place in Paris on the GCT Telescope prototype structure. On-telescope testing will include: the alignment of the camera in the focal plane, in-situ calibration, recording Cherenkov images, and procedural aspects of installing, operating and maintaining the camera as part of the telescope system.

\section{Beyond CHEC-M}\label{GCT-SiPM} 
The second prototype GCT camera (CHEC-S) features similar mechanics and the same back-end electronics as CHEC-M. SiPMs are used instead of MAPMs and an upgrade of ASIC from TARGET 5 to TARGET 7 is present. A revised pre-amplification and shaping scheme is needed to accommodate the SiPM pulse shape and an update to the TARGET modules is implemented. The TARGET 7 ASICs feature improved linearity and dynamic range when compared to TARGET 5. CHEC-S is expected to be fully integrated towards the end of 2015. Beyond CHEC-S the GCT camera will go through one further design iteration before entering the CTA Production phase. During this design iteration the final photosensor choice will be made and a final iteration of the TARGET ASICs will take place.

\section{Summary}

The GCT camera is a low-cost, high-reliability, high-data-quality solution for the SST section of CTA. The camera features full waveform readout with nanosecond sampling across a variable window for all 2048 pixels at a rate of up to 600~Hz. The use of high-channel density TARGET ASICs and the latest advances in multi-pixel photosensor tiles ensure a cost-effective design. The first camera prototype (CHEC-M) has been assembled and is undergoing commissioning in the lab. After full characterisation the camera is scheduled to be installed on a prototype telescope structure in fall 2015. Beyond CHEC-M a second prototype based on SiPMs is under development with the eventual aim of providing 35 cameras to CTA.

\acknowledgments
The research leading to these results has received funding from the People Programme (Marie Curie Actions) of the European Unions Seventh
Framework Programme FP7/2007-2013/ under REA grant agreement n [317446] INFIERI ``INtelligent Fast Interconnected and Efficient Devices
for Frontier Exploitation in Research and Industry``.We would also like to acknowledge the support of the UK Science and Technology Facilities Council (grant ST/K501979/1) and the support from the agencies and organisations listed in this page: http://www.cta-observatory.org.

\end{document}